\documentclass[twocolumn,preprintnumbers,amsmath,amssymb,pra]{revtex4}
\usepackage{amsmath,amssymb,graphicx,epstopdf,bm,soul,upgreek, hyperref,natbib}
\usepackage{graphicx}
\usepackage{color}
\definecolor{OliveGreen}{rgb}{0,0.6,0}
\usepackage{dcolumn}
\usepackage{bm}
\usepackage{amssymb,float}
\usepackage{cancel}
\usepackage{ulem}

\begin{document}
\title{One-way reflection-free exciton-polariton spin filtering channel}%
\author{S. Mandal}\email[Corresponding author:~]{subhaska001@e.ntu.edu.sg}
\author{R. Banerjee}
\author{T.C.H. Liew}\email[Corresponding author:~]{tchliew@gmail.com}

\affiliation{Division of Physics and Applied Physics, School of Physical and Mathematical Sciences, Nanyang Technological University, Singapore 637371, Singapore}

\begin{abstract}
 We consider theoretically exciton-polaritons in a strip of honeycomb lattice with zigzag edges and it is shown that the interplay among the spin-orbit coupling, Zeeman splitting, and an onsite detuning between  sublattices can give rise to a band structure where all the edge states of the system split in energy. Within an energy interval,  one of the spin polarized edge states resides with the gap-less bulk having opposite spin. Being surrounded by opposite spin and the absence of the backward propagating edge state ensures both reflection free and feedback suppressed one-way flow of polaritons with one particular spin in the system. The edge states in this system are more localized than those in the standard topological polariton systems and are fully spin polarized. This paves the way for feedback free spin-selective polariton channels for transferring information in polariton networks. 
\end{abstract}

\maketitle
{\textit{Introduction}---} It has been recognized long ago that one of the crucial ingredients for optical circuits is the isolation of input and output \cite{Keyes_1985}. Equivalently, circuits should operate in a one-way fashion, where feedback in the backward direction is suppressed. As Maxwell's equations are time reversal invariant, the realization of such a requirement is not obvious in many optical systems, and reflections must be considered carefully. With the emergence of topological photonics \cite{Lu_2014,Ozawa_2019}, special techniques were introduced to artificially break time-reversal symmetry (e.g., helical waveguide arrays as in Floquet topological insulators \cite{Lindner_2011}) or exotic material solutions employed at particular wavelengths (e.g., magnetically sensitive gyrotropic materials \cite{Wang_2009}). One of the well-known features of topological photonic lattices is the appearance of chiral edge states. As these propagate in only a certain direction and do not undergo backscattering, it has been natural to expect that such states are relevant for the robust propagation of optical signals.
\begin{figure}[t]
\includegraphics[width=0.5\textwidth]{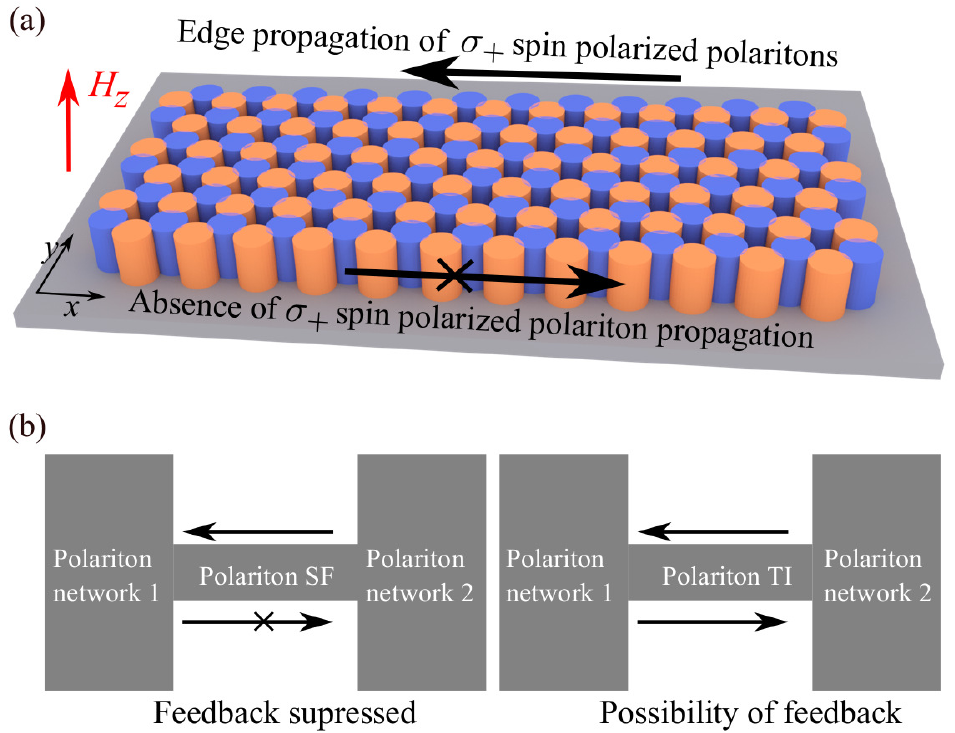}
\caption{(a) Schematic diagram of coupled polariton micropillars arranged in a honeycomb strip where two different sub-lattices are indicated by two colors. The considered system can act as a spin filter by allowing the polaritons with one of the spins in the system to propagate robustly from right to left while the propagation in the opposite direction is restricted. The red arrow indicates the perpendicularly applied magnetic field ($H_z$). (b) Potential application of the polariton spin filter (SF) as a channel that can be used to couple two polariton networks without feedback. The same can not be done using a polariton topological insulator (TI) where there exists a possibility of feedback via an opposite edge state.}
\label{Fig1}
\end{figure}

Instead of breaking time-reversal symmetry directly in a photonic system, an alternative is to couple it to a system in which time-reversal symmetry is broken. In particular, when photons are coupled to excitons in semiconductor micropillars they hybridize to form exciton-polaritons, which inherit a Zeeman splitting due to their excitonic component when placed in a magnetic field \cite{Rahimi-Iman_2011}.  Since such micropillars can be arranged into lattices it has been possible to form topological polariton systems \cite{Klembt_2018}, accounting for the effect of spin-orbit coupling \cite{Karzig_2015,Bardyn_2015,Nalitov_2015}.

Exciton-polaritons also exhibit a significant nonlinearity, which in theory was predicted to itself induce topological behavior \cite{Bardyn_2016}, antichiral behavior \cite{Mandal_2019}, and topological solitons \cite{Kartashov_2016}. Separate from topological physics, nonlinearity was shown to provide low-energy polaritonic switches \cite{Grosso_2014,Dreismann_2016,Lewandowski_2017}  and realize other information processing elements such as transistors \cite{Gao_2012,Ballarini_2013,Lewandowski_2017,Zasedatelev_2019}, routers \cite{Flayac_2013,Marsault_2015,Schmutzler_2015}, and amplifiers \cite{Wertz_2012,Niemietz_2016}. The possible use of topological polariton states at the edges of a strip to couple these elements suffers however from an obvious drawback: such states come in pairs, which propagate in opposite directions on the opposite edges of the strip \cite{Bardyn_2015,Nalitov_2015}. Consequently, there will be an unwanted counter-propagating signal at the opposite edge of the strip leading to potentially detrimental feedback (unless the strip is made very wide, which would lead to other problems in scalability).

Furthermore, it has long been expected that polaritonic devices would make use of the spin degree of freedom \cite{Shelykh_2004}, which can take one of two spin projections on the structure growth axis (denoted $\sigma_\pm$). Individual spin switches \cite{Amo_2010} and gates \cite{Gao_2015} have encoded information in this spin. Unfortunately though, topological polariton states have a specific polarization that mixes the $\sigma_+$ and $\sigma_-$ components \cite{Bardyn_2015,Nalitov_2015}, which would not allow the preservation of individual $\sigma_+$ or $\sigma_-$ spin polarized wavepackets.

In this work, we aim to develop the recent advances in topological polariton physics to introduce a channel for connecting polaritons of a given spin. We aim for the channel to be reasonably thin and allow only one spin to propagate in a specific direction. To meet this aim, we consider a strip of honeycomb polariton lattice, which is composed of two interlocking sublattices, A and B (see Fig. \ref{Fig1}(a)). The sites of the different sublattices are engineered with different onsite energy. Accounting for a magnetic field and spin-orbit coupling, we find that the states at different edges are split in energy and become almost completely spin polarized. Furthermore, an edge state of a given spin can be embedded in an energy range where the only existing states have  opposite spin polarization. As the scattering of polaritons with disorder tends to be spin conserving, this maintains a unidirectional propagating edge state for a specific spin, which is not possible with a regular topological polariton system (see Fig. \ref{Fig1}(b)). A further key difference with existing topological polariton realizations is that the strip can be made remarkably thin (three lattice constants thick) and still maintain its unidirectional spin polarized behavior.

{\textit{Scheme}---} We start by considering a planar semiconductor microcavity containing exciton-polaritons subjected to a resonant excitation, expressed by the driven dissipative Schr$\ddot{o}$dinger equation,
\begin{align}\label{MainEq}
&i\hbar\frac{\partial\psi_{\sigma_\pm}(x,y)}{\partial t}=\left[-\frac{\hbar^2\nabla^2}{2m}+V(x,y)-i\frac{\Gamma}{2}\pm\Delta_z\right]\psi_{\sigma_\pm}(x,y)\nonumber\\
&+\Delta_T\left(i\frac{\partial}{\partial x}\pm\frac{\partial}{\partial y}\right)^2\psi_{\sigma_\mp}(x,y)+F_{\sigma_\pm}(x,y)e^{i(k_0x-\omega_p t)}.
\end{align}
\begin{figure}[t]
\includegraphics[width=0.5\textwidth]{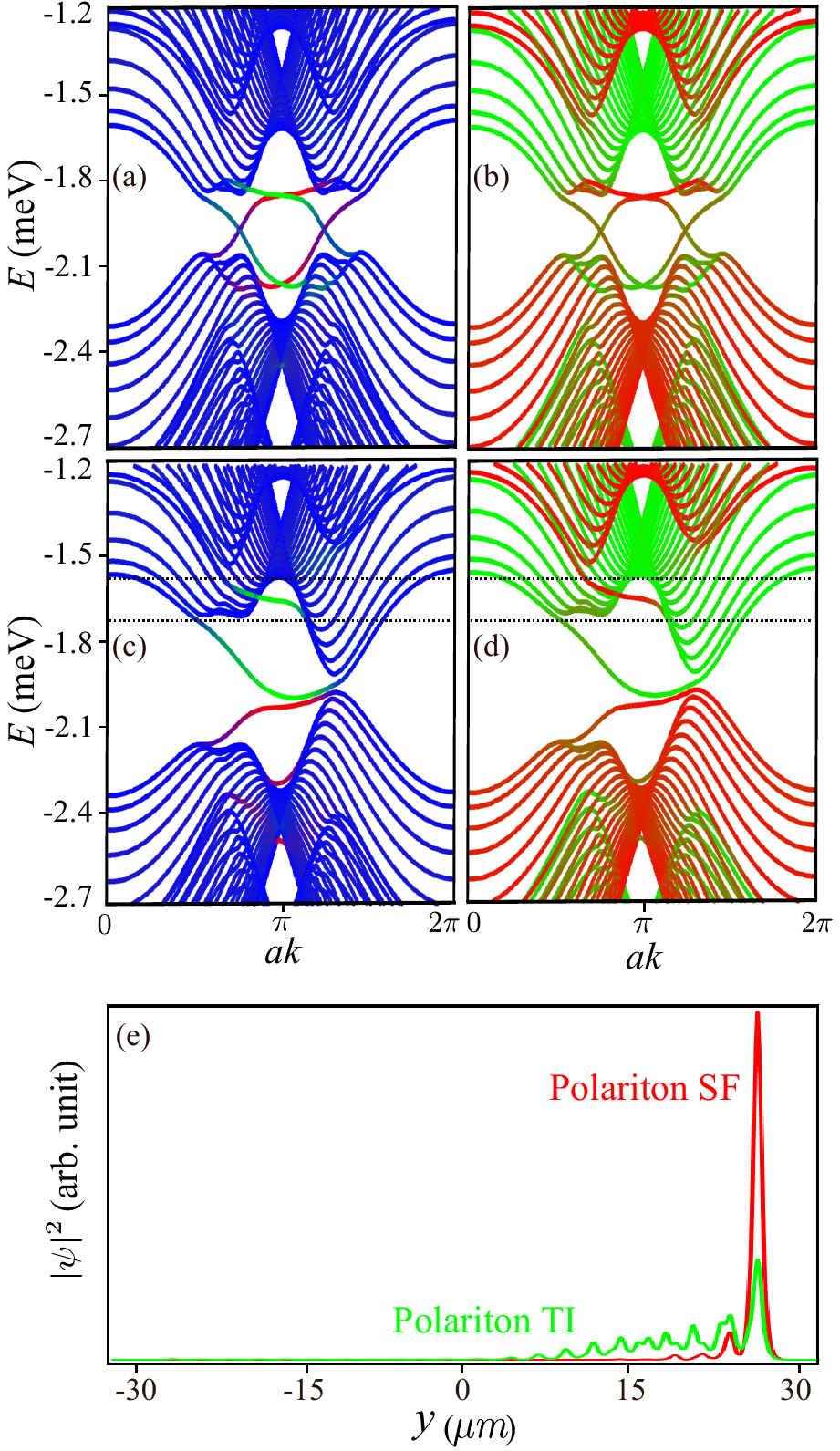}
\caption{Band structure of the system under consideration of onsite energy $\delta=0$ in (a-b) and  $\delta \neq 0$ in (c-d). In (a) and (c) the states residing in bulk are shown in blue whereas red and green correspond to the states located near the two edges. In (b) and (d) the same dispersions with spin contribution are plotted where red corresponds to the states having spin $\sigma_+$ and green corresponds to the states having spin $\sigma_-$. The dashed lines in (c-d) represent the energy window where the edge state is surrounded by the bulk states with opposite spins and also there is no counter propagating edge state. (e) comparison of the localization of the edge states between the polariton SF and polariton TI.}
\label{Fig2}
\end{figure}
Here $\psi_{\sigma_\pm}(x,y)$ are the wave functions for $\sigma_\pm$  spin polarized polaritons; $V(x,y)$ is a potential describing a strip of honeycomb lattice having zigzag edges along the $y$ direction; $\Gamma$ is a uniform decay throughout the sample; $2\Delta_z$ and $\Delta_T$ are  the Zeeman and TE-TM splittings, respectively. $F_{\sigma_\pm}(x,y)$ are the circular polarization components of a resonant excitation with frequency $\omega_p$. Each micropillar is modelled by placing a Gaussian of width $\sigma$ at each site \cite{Kartashov_2017} and the lattice has a periodicity $a$ along the $x$ direction, such that,
\begin{align}\label{potential}
&V(x,y)=V(x+a,y)\nonumber\\
&=-(v_o\pm\delta)\sum_{x_n,y_n} \exp\left[-\frac{(x-x_n)^2+(y-y_n)^2}{\sigma^2}\right].
\end{align}
Here the summation runs over all  sites, $(v_o+\delta)$ is the potential depth corresponding to the A sublattices, and $(v_o-\delta)$ is the potential depth corresponding to the B sublattices.  In principle, the energy detuning, $\delta$, between the A and B sublattices can be induced by locally adjusting the thickness of the cavity or by  non-resonantly pumping one of the sublattices, where the excitonic reservoir will induce a blueshift locally \cite{Galbiati_2012}. Alternatively, micropillars corresponding to the A and B sublattices could be engineered with different diameters (see the supplementary material (SM)). In order to know the characteristics of the system we consider the linear band structure by setting $F_\pm=0$ and $\Gamma=0$ in Eq.~(\ref{MainEq}) and after applying the Bloch theorem the wave function can be written as,
 $ \psi_{\sigma_\pm}(x,y)=u_{\sigma_\pm}(x,y)\exp{(ikx)}$. Here $u_{\sigma_\pm}(x,y)$ are the Bloch wave functions, which are periodic along the $x$ direction and $k$ is the  wave-vector in the $x$ direction. Upon substitution,  Eq.~(\ref{MainEq}) leads to an eigenvalue problem, which can be solved  to obtain the band structure. In Figs.~\ref{Fig2} (a-b) and (c-d) we plot the dispersion without and with the onsite term, respectively. For $\delta=0$, we obtain the band structure corresponding to the polariton topological insulator (TI) (see Fig.~\ref{Fig2} (a)) where the time reversal symmetry is broken by the Zeeman splitting. Here, the bulk modes are gapped and in the band gap robust edge states appear. In this case, the edge states come in counter propagating pairs, meaning that within the band gap of each propagating edge state there is another one  propagating in the opposite direction. The inclusion of $\delta$ breaks the inversion symmetry and  the parameters can be chosen such that the band gap at one of the Dirac points gets closed resulting in a topologically trivial system. The dispersion of the bulk modes is similar to the ones in bilayer graphene with a perpendicular bias \cite{Foa_2016}. Although, in this case the edge states are completely separated in energy (see Fig.~\ref{Fig2} (b)). Note that there is no counter propagating edge mode present in the energy window of interest. We have also plotted the localization of the edge states in Fig.~\ref{Fig2}(e), which shows that the edge states in the spin filter (SF) regime are more localized than the standard polariton TI.

To emphasize the interest of the obtained states, we plot the band structures and colour code them depending upon the contribution of the spins. In the case of a polariton TI (see Fig.~\ref{Fig2} (b)) we find that within the bandgap the edge states have mixed spin. On the other hand in the SF regime (Fig.~\ref{Fig2} (d)) the edge states have almost pure spin and more importantly the top most edge state is  surrounded by only the bulk modes with opposite spin polarization. We can then expect $\sigma_+$ spin polarized polaritons to propagate through the edge mode without backscattering, manifesting  a robust SF. 
\begin{figure}[t]
\includegraphics[width=0.5\textwidth]{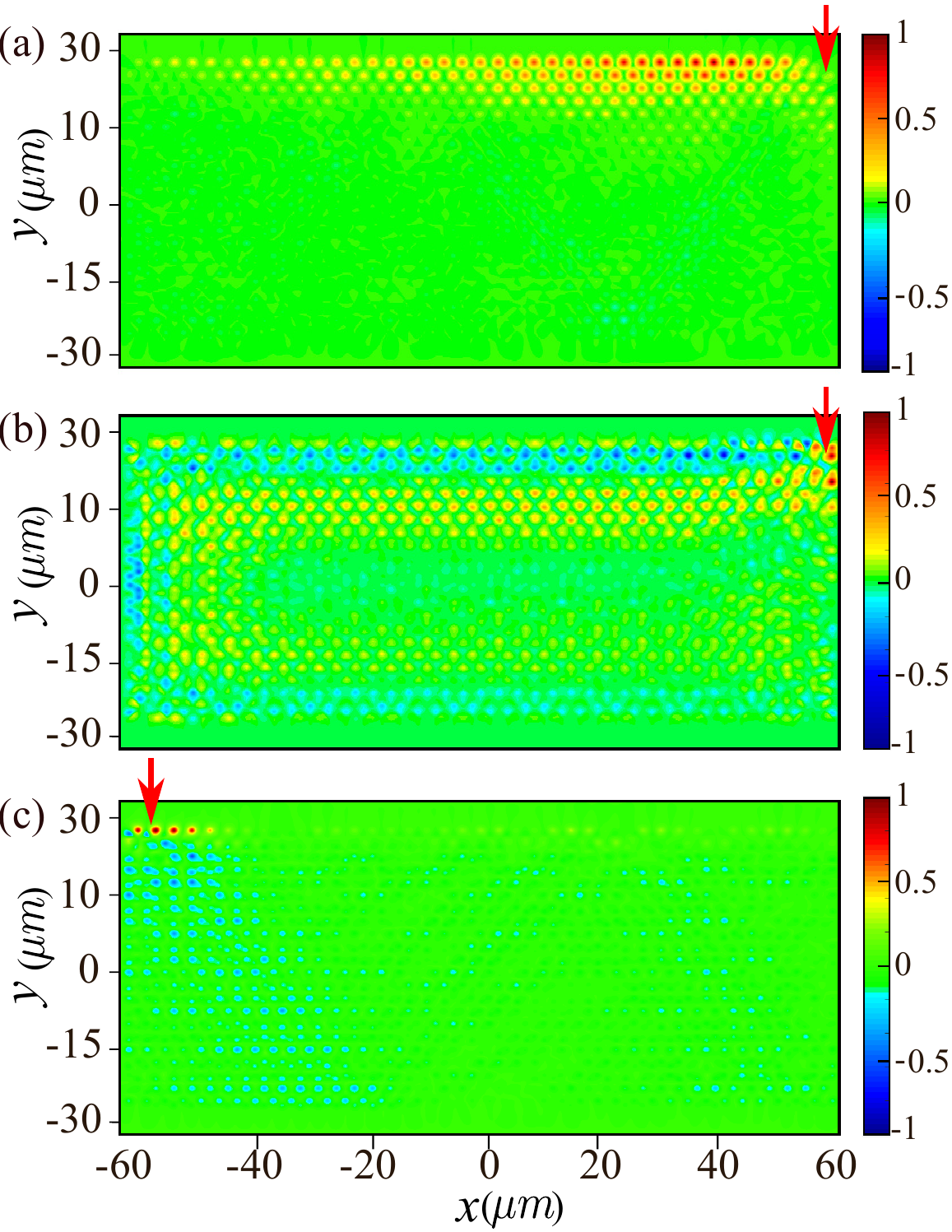}
\caption{Spatial distribution of the spin degree defined in Eq.~\ref{Spin_degree} for the polariton SF is shown in (a), where polaritons having both spins are pumped but only $\sigma_+$ spin polarized polaritons flow one way. The same for polariton TI is shown in (b), where polaritons  with mixed spins flow counter-clock wise through the edges of the sample. (c) Feedback suppression effect where the $\sigma_+$ spin polarized polaritons can not flow from left to right, instead they are converted to $\sigma_-$ spin polarized polaritons and scatter randomly into the bulk. The red arrows indicate the position of the continuous pump. Parameters:  Pump width $=9~ \mu m$; $\Gamma=0.01$ meV; $k_0=2.5~\mu m^{-1}/a$ and  $\omega_p=-1.65$ meV$/\hbar$  are used in (a) and (c); $k_0=2.35~\mu m^{-1}/a$  and $\omega_p=-1.98$ meV$/\hbar$ are used in (b).}
\label{Fig3}
\end{figure}

\begin{figure}[t]
\includegraphics[width=0.5\textwidth]{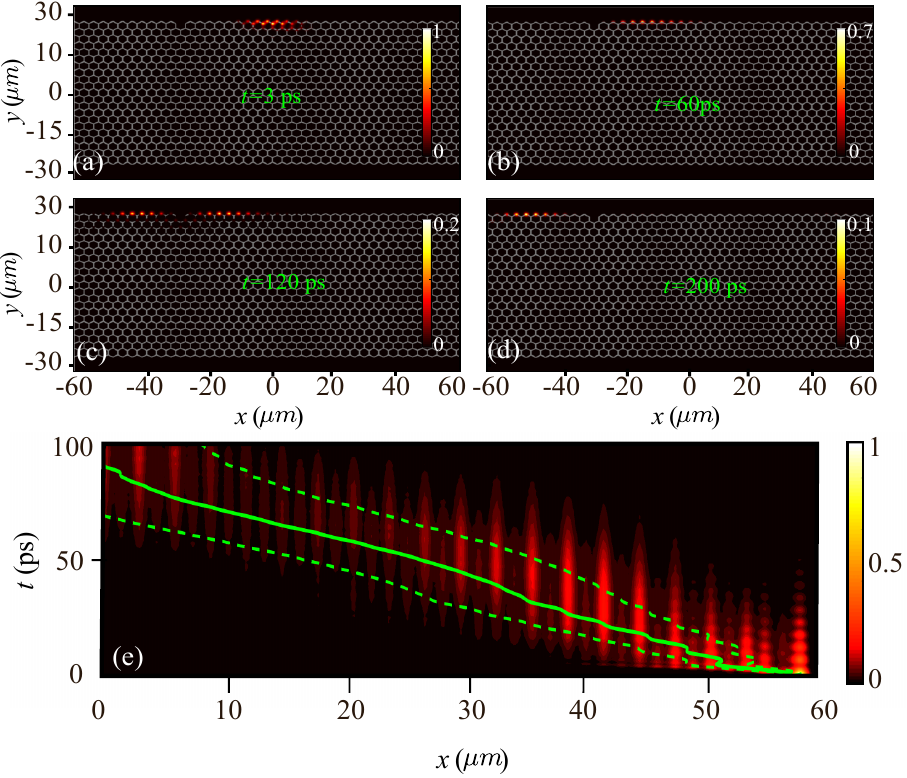}
\caption{(a-d) Propagation of a $\sigma_{+}$ coherent pulse in a honeycomb lattice strip with one of the sites removed. (e) Propagation of the same  pulse in a honeycomb lattice strip with spatial disorder. The colour scale represents the $\sigma_+$ polarized intensity, integrated over the strip width (y-coordinate). The green curve indicates the average position of the wavepacket and the dashed lines the variance. These are defined as $x_m=\int x |\psi_{\sigma_+}|^2dxdy/I_0$ and $x_v^2=\int x^2 |\psi_{\sigma_+}|^2dxdy/I_0-x_m^2$, respectively, where $I_0=\int |\psi_{\sigma_+}|^2dxdy$. The parameters were  the same as in Fig.~\ref{Fig3}(a).}
\label{Fig4}
\end{figure}

Considering typical parameters, we have set: $a=3~\mu$m; $m=2.5\times10^{-5} m_e$ with $m_e$ the free electron mass;   $v_o=16.5$ meV; $\delta=0.54$ meV; $\sigma=0.45~\mu$m; $\Delta_T=0.15$ meV$\mu$m$^2$; and $\Delta_Z=0.18$ meV. This gives an energy interval of  the SF edge state of around 0.12 meV,  which is well above the linewidth of modern samples.  Generally, the SF regime can be found for $\varepsilon_g\simeq 2\Delta_z\le \Delta_T$, where $\varepsilon_g$ is the trivial bandgap induced by the broken inversion symetry (see SM Figs.~ S1- S3) and $\Delta_T$ determines the group velocity of the edge state. It should be stressed that instead of the magnetic field (Zeeman splitting), the nonresonant excitation can itself break the symmetry between the two spins \cite{Bleu_2016,Sigurdsson_2019}. For the rest of the paper we will focus on the top most edge state and although here only $\sigma_+$ spin polarized polaritons are filtered, the same can be done for the $\sigma_-$ spin polarized polaritons (see SM). 

{\textit{Spin filtering}---} To validate our claim we excite the top most edge state resonantly with a continuous linearly polarized Gaussian shaped pump by setting $F_{\sigma_+}=F_{\sigma_-}$. The spin polarization degree is defined as,   
\begin{equation}\label{Spin_degree}
S(x,y)=\frac{|\psi_{\sigma_+}(x,y)|^2-|\psi_{\sigma_-}(x,y)|^2}{|\psi_{\sigma_+}(x,y)|^2+|\psi_{\sigma_-}(x,y)|^2}.
\end{equation}
The spatial distribution of $S(x,y)$ is plotted in Fig.~\ref{Fig3}(a) in the SF regime,  which shows that although we have excited both spin states, only $\sigma_+$ spin polarized polaritons populate the top edge and propagate over $100~\mu m$ from the spot of excitation. Due to the finite lifetime the intensity decreases with the increase of propagation distance. $\sigma_-$ spin polarized polaritons are instead scattered randomly into the bulk. To demonstrate that there is no state of $\sigma_+$ spin polarized polaritons to propagate in the reverse direction, we consider injecting both spins from the opposite corner of the top edge (see Fig.~\ref{Fig3}(c)). We observe that the $\sigma_+$ spin polarized polaritons can not propagate from left to right of the system; instead  they are converted to $\sigma_-$  polaritons and scatter into the bulk. This confirms that the system can act as a feedback free unidirectional spin filter. On the other hand, in a polariton TI the polaritons with mixed spins rotate through the edges of the sample in anticlockwise direction (See Fig.~\ref{Fig3}(b)). Although the exact contribution of the spins in $S(x,y)$ can be changed slightly depending upon the state that is being excited, there is always a significant contribution from both the spins.

 In all our calculations, we have taken a lifetime around 60 ps, close to that reported in \cite{Klembt_2018,Galbiati_2012}. The SF would also work with shorter lifetime, although the propagation distance would decrease (see SM, Fig.~ S3). Higher lifetimes such as those reported in \cite{Sun_2017} are thus the ideal.

{\textit{Pulse propagation around defect}---}
Fig. \ref{Fig4} (a-d) consider the propagation of a $\sigma_+$ polarized pulse (obtained by multiplying the coherent excitation with a Gaussian in time) in the case where a micropillar is removed from the top edge.  We observe that near the defect the pulse spreads a little but  does not backscatter and after some time the pulse passes the defect. This further proves the robustness of the edge state.

\begin{figure}[t]
\includegraphics[width=0.5\textwidth]{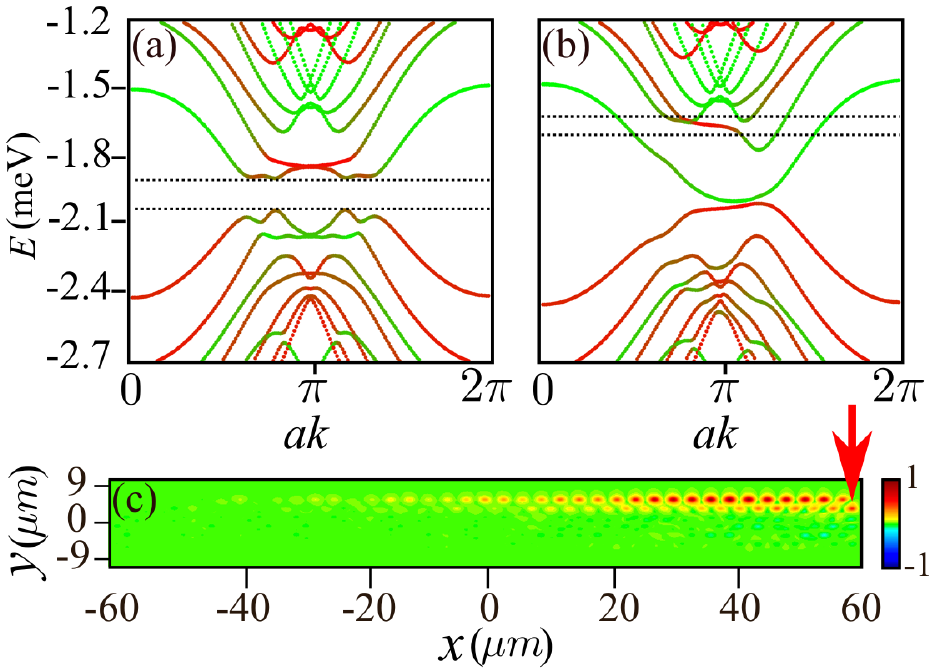}
\caption{Band structure corresponding to a very short strip consisting of three unit cells along the $y$ direction for TI shown in (a) and the same for SF shown in (b). The edge states in the topological gap corresponding to the TI disappear. (c) Spatial distribution of the spin degree for SF. The red arrow indicates the position of continuous pump. Parameters were kept the same as those in  Fig.\ref{Fig3}(a).}
\label{Fig5}
\end{figure}

{\textit{Propagation in the presence of disorder}---} In real systems, disorder is present throughout. Fig.~\ref{Fig4} (e) shows the propagation of a $\sigma_+$ polarized pulse in the presence of a continuous disorder potential, which is characterized by a root mean squared amplitude of $30\mu$eV and a correlation length of $3\mu$m (corresponding to typical experimental values \cite{Baboux_2016}). Again, no backscattering of the $\sigma_+$ polarized wavepacket is observed.
 
{\textit{Strip width}---}
In addition to supporting one-way propagation, a channel for connecting information processing elements should not be too wide, so as not to limit the footprint area of future systems. In Figs.~\ref{Fig5} (a-b) the band structure for a narrow strip having three unit cells along the $y$ direction is plotted for a polariton TI and SF, respectively. Due to the lack of bulk states, the well spread edge states in the polariton TI (see Fig.~\ref{Fig2}(e)) vanish while comparatively more localized edge states in the polariton SF remain unaffected resulting in $\sigma_+$ spin  propagation as shown in Fig.~\ref{Fig5} (c). 

{\textit{Discussion}---}
Modern electronic devices and integrated circuits are limited by heating and time delay of communication between components. Optical devices seem promising candidates for fast interconnections and higher fidelity may reduce error correction overhead, resulting in less heating \cite{Miller_2017}. But optics needs to operate at lower threshold to be relevant, for which nonlinear optical systems are essential. Along with the nonlinearity, the spin degree  freedom of exciton-polaritons make them extremely promising for nonlinear spintronics devices. The coupling between the exciton polariton condensates in lattices have been demonstrated but this coupling is always bidirectional and consequently, there will be unwanted feedback. Although the proposed spin filter is not itself a device it can be an essential component of  complete future circuits or networks. 

{\textit{Conclusion}---} We have proposed theoretically a scheme to obtain an optical spin filter based on  exciton-polaritons. The naturally present spin orbit interaction along with Zeeman splitting and an onsite detuning between  sublattices in a polariton honeycomb lattice gives rise to a band structure where all the edge states of the system split in energy and there exists an energy window where one of the edge states with pure spin is surrounded by  bulk states with opposite spin. Since there is no state with the same spin to backscatter to, polaritons having one of the spins in the system can propagate robustly through the edge state and thus  spin filtering is achieved. The absence of the pairing backward propagating edge state reduces feedback in the system. Furthermore, the edge state in the polariton SF is more localized than those in a polariton TI, which allows to operate with very narrow channels. In contrast, regular polariton TI systems would require much wider strips/lattices to support robust propagation and would cause mixing between spin polarized states.

{\textit{Acknowledgement}---} The work was supported by the Ministry of Education, Singapore (grant nos. MOE2018-T2-02-068 and MOE2018-T3-1-002).

\end{document}


\title{Supplemental material for One-way reflection-free exciton-polariton spin filtering channel}%
\author{S. Mandal}\email[Corresponding author:~]{subhaska001@e.ntu.edu.sg}
\author{R. Banerjee}
\author{T.C.H. Liew}\email[Corresponding author:~]{tchliew@gmail.com}

\affiliation{Division of Physics and Applied Physics, School of Physical and Mathematical Sciences, Nanyang Technological University, Singapore 637371, Singapore}
\maketitle
\section{Band structure of the spin filter using micropillars of different diameters}
\begin{figure}[H]
\includegraphics[width=0.5\textwidth]{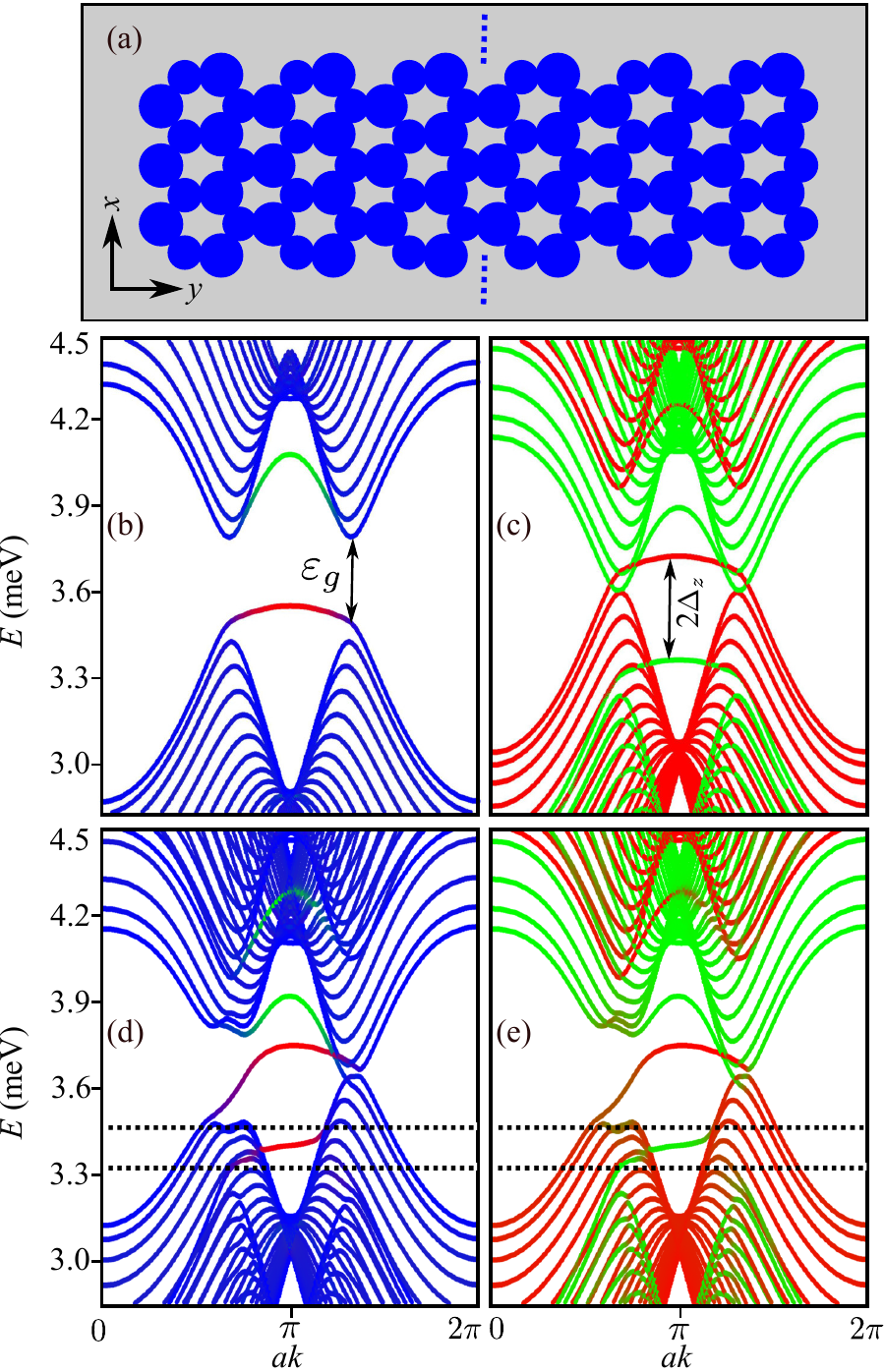}
\caption{(a) Schematic diagram of a honeycomb lattice having zigzag edges consisting of two different types of micropillars having different diameters. Band structure of the system under consideration for $\Delta_z=\Delta_T=0$ in (b);  for $\Delta_z = 0.18$ meV and $\Delta_T=0$ meV$\mu$m$^2$ in (c); and for $\Delta_z = 0.18$ meV and $\Delta_T=0.1$ meV$\mu$m$^2$ in (d-e). In (b) and (d) the bulk modes are represented in blue whereas green and red represent the states located at the two edges. In (c) and (e) red and green represent the $\sigma_+$ and $\sigma_-$ spin polarized modes, respectively. The mass of the polaritons is 2.5 $\times$ $10^{-5}m_e$ with $m_e$ the free electron mass. }
\label{FigS1}
\end{figure}

\begin{figure}[t]
\centering
\includegraphics[width=0.5\textwidth]{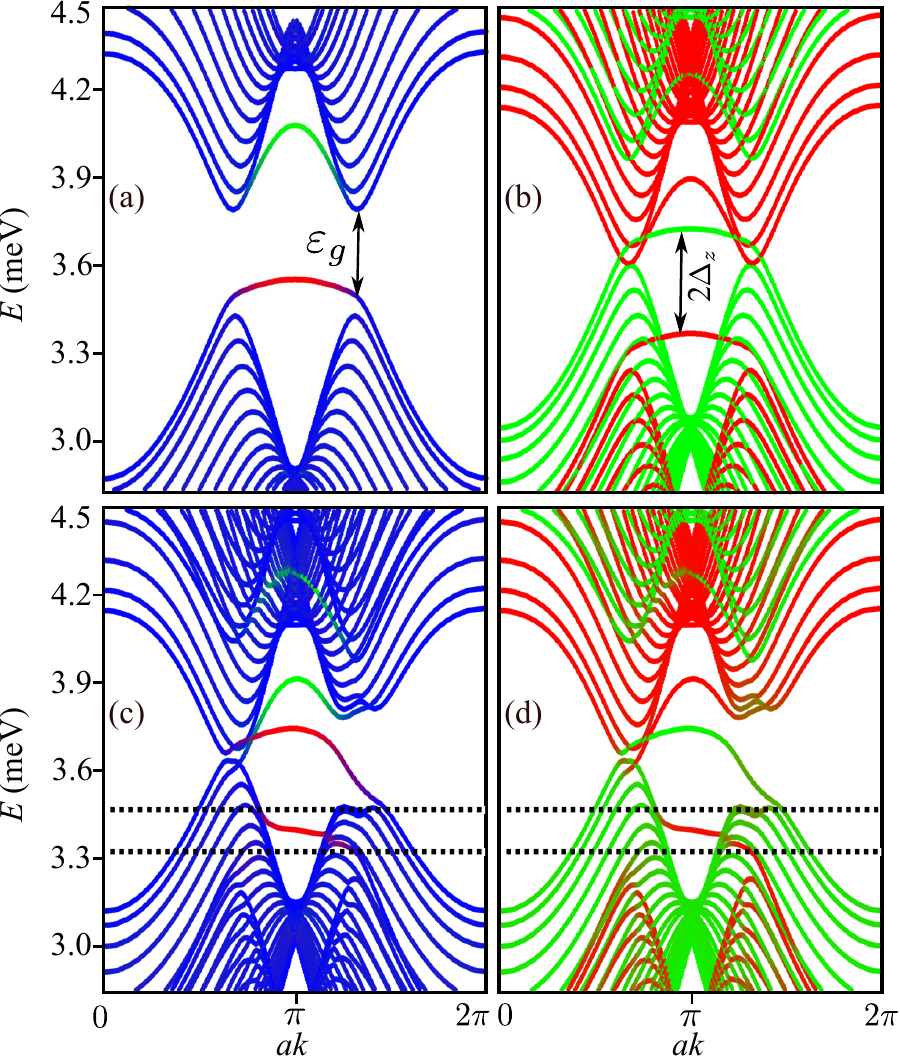}
\caption{Band structure of the system shown in Fig.~\ref{FigS1}(a) for $\Delta_z=\Delta_T=0$ in (a);  for $\Delta_z =- 0.18$ meV and $\Delta_T=0$ meV$\mu$m$^2$ in (b); and for $\Delta_z = -0.18$ meV and $\Delta_T=0.1$ meV$\mu$m$^2$ in (c-d). The colour codings are the same as those in Fig.~\ref{FigS1}. In this case the SF works for $\sigma_+$ spin polarized polaritons.}
\label{FigS1_2}
\end{figure}

In this section, we consider micropillars of two different diameters to construct the honeycomb lattice. In the considered system two types of micropilars have different mode localization energies due to their different diameters, which automatically gives rise to the onsite detuning, $\delta$. In Fig.~\ref{FigS1}(a) a schematic diagram of such a system of micropillars arranged in a honeycomb lattice is considered, which is  infinite along the $x$ direction and finite along the $y$ direction having the zigzag edges. The band structure is calculated the same way as done in the main text. The diameters of the micropillars are considered to be $2.4~\mu$m and $1.9~\mu$m \cite{Klembt_2018}, which represent the A and B sublattices, respectively, and the depth of the micropillars are the same and taken 16.5 meV. In order to realize the coupling between the two neighboring sites, the micropillars have an overlap between them such that the unit cell size along the $x$ direction is 3 $\mu$m. In Fig.~\ref{FigS1}(b) the band structure of the system is plotted without taking into account the Zeeman splitting, $2\Delta_z$, and the TE-TM splitting, $\Delta_T$. There is a trivial bandgap, $\varepsilon_g$, which is due to the breaking of the inversion symmetry. Since, the edges of the lattice are composed of different types of micropillars, the edge states located at the two edges of the lattice are no longer symmetric. There is no term ($\Delta_z=0$) to break the symmetry between the $\sigma_{+}$ and  $\sigma_{-}$ spin polarized polaritons and as a result the band structure is degenerate for both spins. The effect of $\Delta_z$ is shown in Fig.~\ref{FigS1}(c), which lifts the degeneracy between the two spins and the band structure of the $\sigma_+$ and $\sigma_-$ polarized polaritons gets shifted in energy by  $2\Delta_z$. Figs.~\ref{FigS1}(d-e) show the SF band structure taking into account both $\Delta_z$ and $\Delta_T$ terms. The parameters are chosen such that the $\sigma_-$ spin polarized edge state is surrounded by the $\sigma_+$ spin polarized bulk modes and hence one can show in the same way as done in the main text that the $\sigma_-$ spin polarized polaritons will propagate robustly along the lower edge from left to right (determined by the group velocity of the edge state which is positive in this case)  and the system will act as a SF where $\sigma_-$ spin polarized polaritons will be filtered out. It should be noted that, by reversing the sign  of the magnetic field (which changes $\Delta_z$ to $-\Delta_z$ ) one can easily change the spin polarization of the SF edge state (see Fig.~\ref{FigS1_2}).

\section{Effect of TE-TM splitting on the band structure}
\begin{figure}[H]
 \centering
\includegraphics[width=0.49\textwidth]{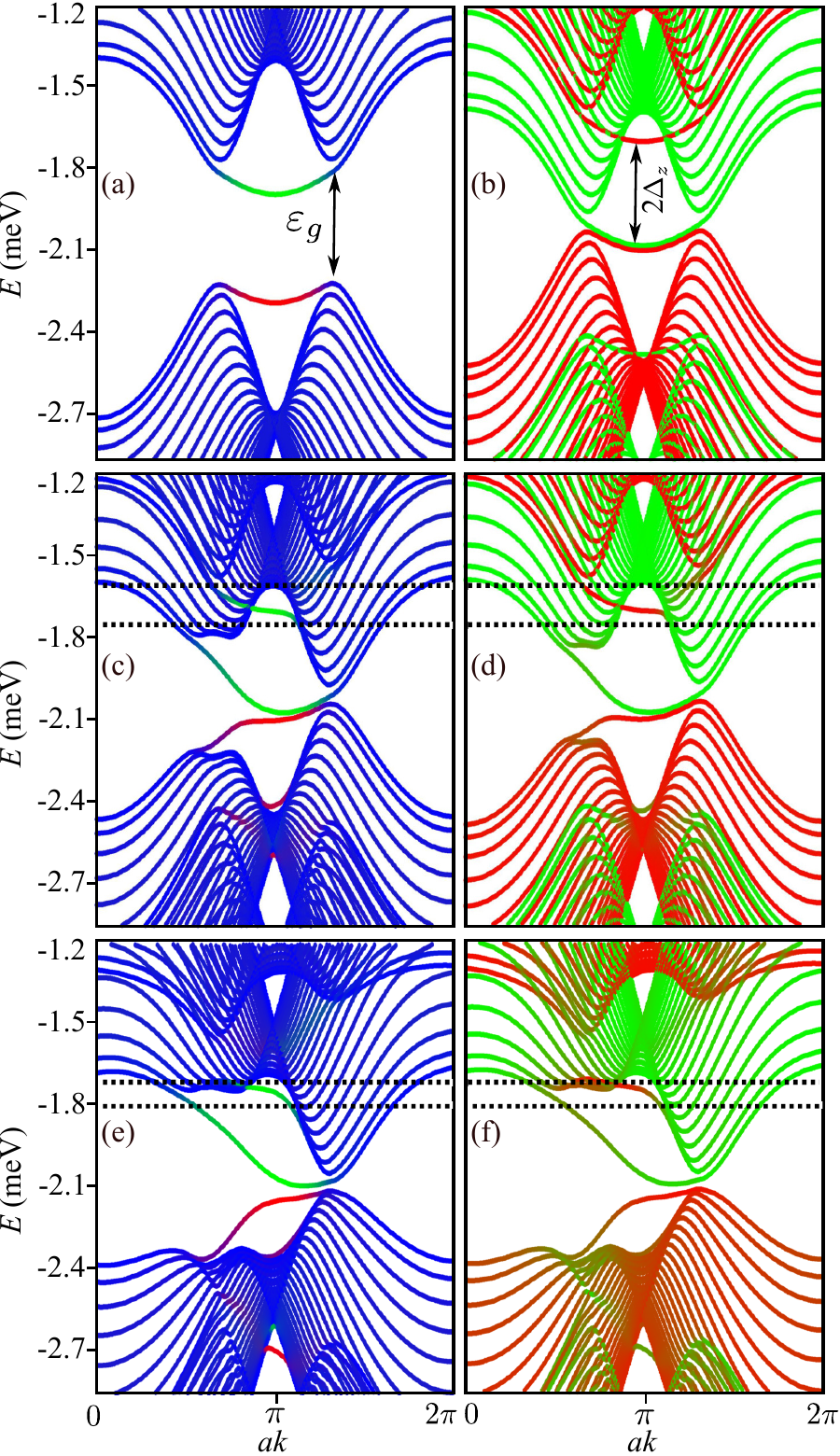}
\caption{Band structure of the system under consideration for $\Delta_z=\Delta_T=0$ in (a),  for $\Delta_z = 0.18$ meV and $\Delta_T=0$ meV$\mu$m$^2$ in (b), for $\Delta_z = 0.18$ meV and $\Delta_T=0.09$ meV$\mu$m$^2$ in (c-d), and for $\Delta_z = 0.18$ meV and $\Delta_T=2.4$ meV$\mu$m$^2$ in (e-f). The colour codings are the same as those in Fig.~\ref{FigS1}.}
\label{FigS2}
\end{figure}

In this section we obtain the band structure by solving Eqs. (1-3) of the main text corresponding to different $\Delta_T$. In Fig.~\ref{FigS2}(a) the band structure of the system without $\Delta_z$ and $\Delta_T$ is plotted which shows the trivial bandgap similar to Fig.~\ref{FigS1}(b). Fig.~\ref{FigS2}(b) shows the effect of $\Delta_z$  which shifts the band structure for $\sigma_\pm$ spin polarized polaritons in energy. Figs.~\ref{FigS2}(c-d) the band structures corresponding to $\Delta_T=0.09$ meV$\mu$m$^2$ are plotted which shows that the SF regime can still exist even for a lower value (than the one used in the main text) of TE-TM splitting. However, the band structure starts to deviate from the SF regime for larger values of TE-TM splitting, $\Delta_T=0.24$ meV$\mu$m$^2$, as shown in \ref{FigS2}(e-f). The SF edge state starts to be flat and also two edge states start to appear within the same energy interval. To conclude, there is a great degree of freedom in choosing the parameters as long as $\varepsilon_g\simeq 2\Delta_z \ge  \Delta_T$. 

\section{Effect of the  polariton  lifetime on the propagation distance}
In this section, the propagation distance for different polariton lifetime is plotted using Eqs. (1-2) and (4) in the main text. For polaritons with lifetime 30 ps the propagation distance is about 60 $\mu$m. As expected, the propagation distance increases with polariton lifetime and the propagation distance reaches around 100 $\mu$m for 50 ps. 
\begin{figure}[H]
\centering
\includegraphics[width=0.5\textwidth]{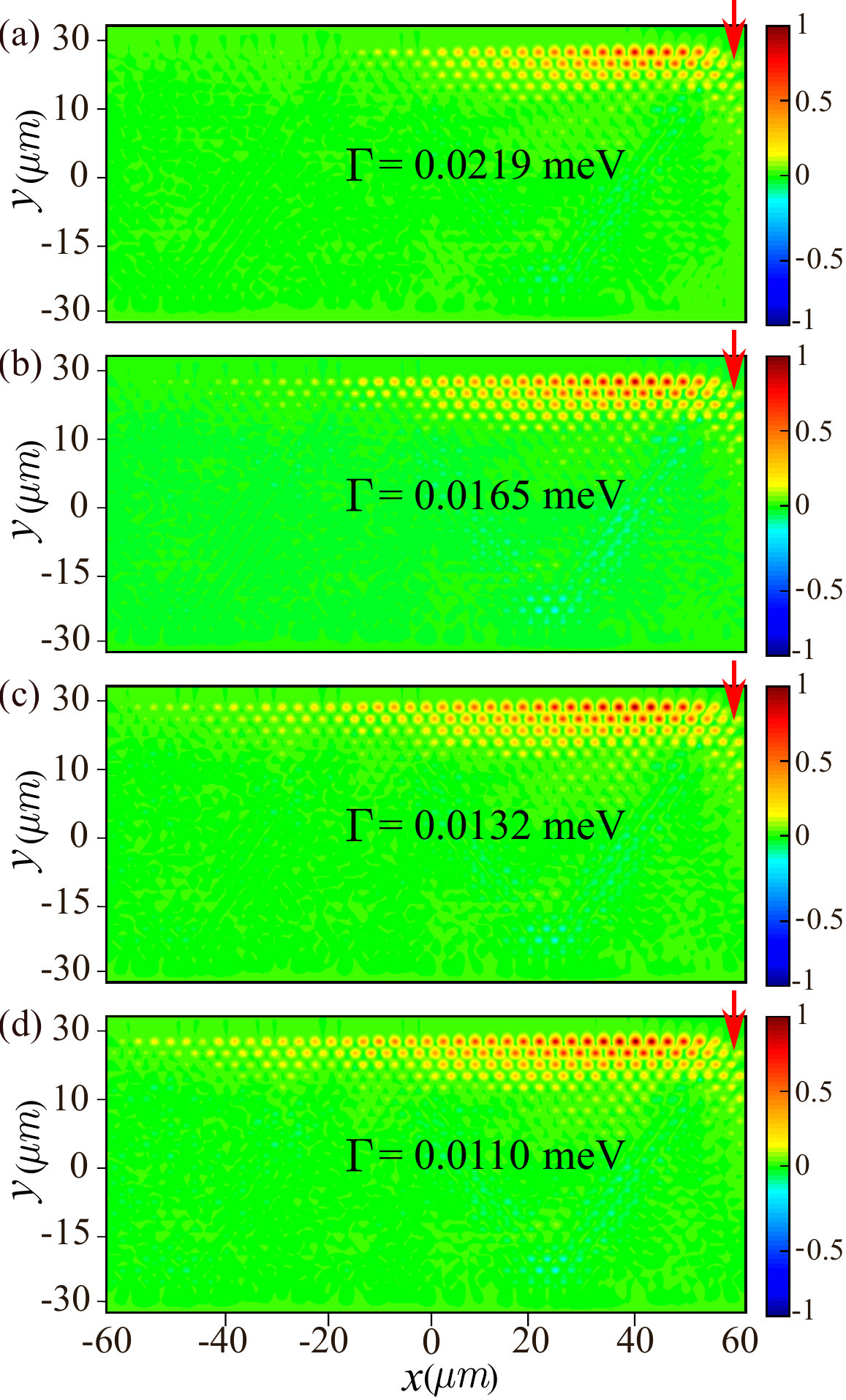}
\caption{Spin propagation in real space for different values of $\Gamma$ which corresponds to the polariton lifetime of 30 ps in (a), 40 ps in (b), 50 ps in (c), and 60 ps in (d). The red arrows indicate the position of the continuous pump. All other parameters were kept the same as those in Fig. 3(a) in the main text.}
\label{FigS3}
\end{figure}